\documentclass[12pt, prd, showpacs]{revtex4}
\usepackage{amssymb}
\usepackage{amsmath}

\input{tcilatex}

\begin{document}

\title{Ultra-high energy head-on collisions without horizons or naked
singularities: general approach}
\author{O. B. Zaslavskii}
\affiliation{Department of Physics and Technology, Kharkov V.N. Karazin National
University, 4 Svoboda Square, Kharkov, 61077, Ukraine}
\email{zaslav@ukr.net }

\begin{abstract}
Recently, alternatives to the Ba\~{n}ados, Silk and West (BSW) effect were
proposed which are (i) due to the existence of naked singularities instead
of the horizon, (ii) require neither horizon nor naked singularity. We
reveal the main features of such alternatives in a model-independent way.
The metric should be close to that of the extremal black hole but the
horizon should not form. Then, one can gain unbound the energy $E_{c.m.}$ in
the centre of mass frame due to head-on collision of particles near the
would-be horizon. The energy measured at infinity can also be unbound. If
instead of particles self-gravitating shells collide, the underlying reason
leading to unbound $E_{c.m.}$ is the same.
\end{abstract}

\maketitle

\section{Introduction}

In recent years, interest to high-energy collisions of particles in strong
gravitational field increased significantly after observation made by Ba\~{n}%
ados, Silk and West (hereafter, BSW). They noticed that if two particles
collide in the vicinity of the Kerr black hole, the energy in the centre of
mass (CM) frame $E_{c.m.}$ may grow unbound \cite{ban}. Later on, it was
shown that the BSW effect is due to the general properties of the black hole
horizon \cite{prd}, so in this sense, the effect has an universal character.
Meanwhile, there are some difficulties in astrophysical realization and
observation of the BSW effect.\ The first one consists in that one of
particles should have fine-tuned relation between the energy and angular
momentum (so-called critical particle). The second one is due to the fact
that enormous $E_{c.m.}$ lead to relatively modest energies measured at
infinity \cite{p} - \cite{z}.

Meanwhile, alternative mechanisms of getting ultra-high energies in the CM
frame also exist. One of them comes back to the works \cite{psk}, \cite{ps}
where unbound $E_{c.m.}$ were also obtained. The crucial difference between 
\cite{psk}, \cite{ps} and \cite{ban} consists in the type of trajectories of
colliding particles. In the BSW effect, both particles approach the horizon.
In \cite{psk}, \cite{ps}, they move in opposite directions, so one of
particles has to move away from the horizon that is rather difficult to
realize for a black (not white) hole. However, there is no need for
fine-tuning parameters in this case \cite{cqg}.

Another mechanism does not require the presence of the horizon at all. The
following scenario in the background of naked Kerr \cite{kerr} and
Reissner-Nordstr\"{o}m (RN) \cite{rnn} metrics were considered. The first
particle reflects from an infinite potential barrier and collides with the
second one. It turned out that $E_{c.m.}\,$\ can be made as large as one
likes provided the parameters of a metric are close to the threshold of
forming an extremal black hole, so the charge $Q=M(1+\varepsilon )$ or $%
J=M^{2}(1+\varepsilon )$ where $M$ is the mass, $Q$ is the charge, $J$ is
the angular momentum, $\varepsilon \ll 1$. Quite recently, it was shown that
unbound $E_{c.m.}\,$are still possible even if both horizons and naked
singularities are absent \cite{nn1}. Such a scenario was further considered
in detail for a particular example of colliding spherical dust shells \cite%
{4}.

The aim of the present work is to extend the approach of \cite{nn1} from the
spherically symmetric case to axially symmetric rotating configurations. We
draw attention that scenario with neither horizons nor naked singularities
has a general character. It does not require the knowledge of the whole
dynamics. We consider generic configurations with matter and reveal the main
features of the effect in a model-independent way. The key ingredients are
(i) the metric on the threshold of forming the extremal horizon, (ii)
head-on collision that is similar to \cite{psk}, \cite{ps} but without
horizons. Point (i) is a feature typical of quasiblack holes (QBH) \cite{qbh}
for which the ultra-high energy collisions are also possible \cite{acqbh}.
However, the mechanism of getting large $E_{c.m.}\,$\ in both cases is
essentially different (see below). Thus instead of considering particular
metrics, trajectories or models \cite{kerr}, \cite{rnn}, \cite{4}, \cite{ax}
we reveal and discuss underlying factors that ensure the existence of the
effect.

All features considered in the present paper imply that although there is no
horizon as such, there exists a time-like surface "close" to it. There, the
value of the lapse function becomes small although not exactly equal to
zero. Meanwhile, there is also another type of the high energy process which
is due to the ergosphere, not the horizon \cite{ergo}, \cite{myergo}. We do
not discuss it here.

\section{General formalism}

Let us consider the metric

\begin{equation}
ds^{2}=-N^{2}dt^{2}+g_{\phi }(d\phi -\omega dt)^{2}+\frac{dr^{2}}{A}%
+g_{\theta }d\theta ^{2}\text{,}  \label{met}
\end{equation}%
where the coefficients do not depend on $t$ and $\phi $. We use units in
which fundamental constants $G=c=$%
%TCIMACRO{\U{127}}%
%BeginExpansion
h{\hskip-.2em}\llap{\protect\rule[1.1ex]{.325em}{.1ex}}{\hskip.2em}%
%EndExpansion
$=1.$

Equations of motion for a particle with the mass $m$ in the backrground (\ref%
{met})\ read%
\begin{equation}
m\dot{t}=\frac{X}{N^{2}}\text{,}  \label{t}
\end{equation}%
\begin{equation}
m\dot{\phi}=\frac{L}{g_{\phi }}+\frac{\omega X}{N^{2}}\text{,}  \label{fi}
\end{equation}%
where dot denotes derivative with respect to the proper time. Here, 
\begin{equation}
X=E-\omega L\text{,}  \label{x}
\end{equation}%
$E=-mu_{0}$, $L=mu_{\phi }$.

From the normalization condition it follows that%
\begin{equation}
m\dot{r}=\pm \frac{\sqrt{A}}{N}Z\text{,}  \label{r}
\end{equation}%
\begin{equation}
Z=\sqrt{X^{2}-N^{2}\left( \frac{L^{2}}{g_{\phi }}+g_{\theta }(p^{\theta
})^{2}+m^{2}\right) },  \label{z}
\end{equation}%
where $p_{\theta }=m\dot{\theta}$.

For geodesic motion, $E$ and $L$ are conserved and have the meaning of the
energy and angular momentum, respectively. However, the equations (\ref{t})
- (\ref{z}) are valid even if $E$ and $L$ are not conserved.

The key quantity of interest is the energy in the CM frame $E_{c.m.}$ If two
particles collide in some point, it can be defined in this point by analogy
with the standard relation for one particle. For two particles with masses $%
m_{1}$ and $m_{2}$ and four-velocities $u_{1}^{\mu }$ and $u_{2}^{\mu },$
the energy $E_{c.m.}$ at the collision event is the norm of their total
four-momentum, 
\begin{equation}
E_{c.m.}^{2}=-(p_{1}^{\mu }+p_{2}^{\mu })(p_{1\mu }+p_{2\mu
})=m_{1}^{2}+m_{2}^{2}+2m_{1}m_{2}\gamma  \label{cm}
\end{equation}%
where 
\begin{equation}
\gamma =-u_{1\mu }u_{2}^{\mu }  \label{gamma}
\end{equation}%
is the relative Lorentz factor.

Then, by direction substitution into (\ref{gamma}), one can find that

\begin{equation}
m_{1}m_{2}\gamma =\frac{X_{1}X_{2}+\delta Z_{1}Z_{2}}{N^{2}}-\frac{L_{1}L_{2}%
}{g_{\phi }}-g_{\theta }p_{1}^{\theta }p_{2}^{\theta }.  \label{ga}
\end{equation}%
where $\delta =-1$ for particles moving in the same radial direction before
collision and $\delta =+1$ otherwise. From now on, we consider scenarios for
which $\delta =+1$ in (\ref{ga}) (head-on collisions).

\section{Metric on threshold of forming the extremal horizon}

In what follows we assume that (i) $N^{2}>0$ everywhere, (ii) for some value 
$r=r_{0}$, it can be made as small as one likes, (iii) collision occurs in
the point $r=r_{0}$. With these assumptions, a natural representation is

\begin{equation}
N^{2}(r,\theta )=B(r,\theta )(r-r_{+})(r-r_{-})  \label{nb}
\end{equation}%
with $B(r_{0},\theta )>0$ separated from zero. It follows from (i) that both
roots are complex and mutually conjugate.\ It is convenient to introduce the
new parameter $\varepsilon $ and write $r_{\pm }=r_{0}\pm ir_{0}\varepsilon $%
, so%
\begin{equation}
N^{2}=B(r,\theta )[(r-r_{0})^{2}+r_{0}^{2}\varepsilon ^{2}]  \label{e}
\end{equation}

Then, requirement (ii) leads to $\varepsilon \ll 1$. Then, near the point of
collision $r-r_{0}=r_{0}O(\varepsilon )$, $N^{2}=O(\varepsilon ^{2}).$
Correspondingly, 
\begin{equation}
\gamma =O(\varepsilon ^{-2})  \label{la}
\end{equation}%
and can be made as large as one likes. In doing so, the metric is perfectly
regular in the vicinity of $r_{0}$. It is seen from above consideration that
under continuous change of the parameter $\varepsilon $, $\ $the system
passes through the state of the extremal black hole when $\varepsilon =0$.
For $\varepsilon \ll 1$. eq. (\ref{cm}) gives us now 
\begin{equation}
E_{c.m.}^{2}\approx \frac{4X_{1}(r_{0},\theta )X_{2}(r_{0},\theta )}{%
B(r_{0},\theta )r_{0}^{2}\varepsilon ^{2}}.  \label{ed}
\end{equation}

If one tries to repeat the procedure for the nonextremal would-be horizons,
one is led to take real distinct roots in (\ref{nb}). However, this is
inconsistent with assumption (i) since $N^{2}$ changes the sign when $r$
passes through $r_{-}$ and $r_{+}$. Therefore, the effect under
consideration is impossible. It is worth reminding that, by contrast, the
BSW effect for nonextremal horizons is possible \cite{gp}, \cite{prd}.

\subsection{Example: the Reissner-Nordstr\"{o}m metric}

To illustrate the general situation, one can compare it to the previous
results for the RN metric \cite{rnn}. In this case, 
\begin{equation}
N^{2}=1-\frac{2M}{r}+\frac{Q^{2}}{r^{2}}.  \label{nq}
\end{equation}

Let for simplicity two particles have the same mass $m$ and collision occurs
in the point where $N^{2}$ reaches the minimum value. Then, according to eq.
16 of \cite{rnn}, 
\begin{equation}
E_{c.m.}^{2}=\frac{4m^{2}}{1-\frac{M^{2}}{Q^{2}}}\text{.}
\end{equation}

Thus, unbound $E_{c.m.}^{2}$ correspond to $Q\rightarrow M$, so the metric
looks "almost" like the extremal RN black hole. There is a naked singularity
inside at $r=0$ due to the last term in (\ref{nq}). This terms is also
responsible for repulsion and reflection of the first particle. However, in
a general case one cam imagine some distribution of matter which smooths the
singularity.

\section{Mechanism of collision}

Let particle 1 pass over $r=r_{0}$ and bounce back at some $r_{1}<r_{0}$.
Then, it collides with particle 2 that moves from the outside region. If the
point of collision is adjusted to be at $r_{0}$ or in its immediate
vicinity, we obtain $E_{c.m.}\sim \varepsilon ^{-2}$ in accordance with (\ref%
{cm}), (\ref{la}). To make particle 1 to reflect at $r=r_{1}$, some
potential barrier should exist at $r<r_{0}$. In some cases it is infinite
like in the RN case \cite{rnn}. Then, the effect under discussion reveals
itself for any Killing energies. However, even if the potential barrier is
of some finite height, the effect of unbound $E_{c.m.}$ persists, although
with the restriction on the permitted range of energies $E$. As $%
N\rightarrow 0$ near $r_{0}$ but $N=O(1)$ inside, it is clear that such a
barrier does exist.

To some extent, the situation resembles that for quasiblack holes (QBH) in
that the horizon is "almost" formed but does not form. However, there are
crucial differences. We do not require $N$ to be small everywhere inside in
contrast to the QBH case \cite{qbh}. And, now particles move in the opposite
direction before collision whereas it was assumed in \cite{acqbh} that they
move in the same direction before collision, the BSW effect being due to the
difference in the energy scales outside and inside the quasihorizon.

\section{Time before collision}

It is instructive to evaluate the time needed for collision and compare it
with the corresponding time in the case of the BSW process. As is known, if
we want the BSW effect to occur, we should choose one of particle to be
'critical" (with fine-tuned parameters). And, for such a particle the proper
time required to reach the horizon diverges \cite{ted}, \cite{gp}, \cite{prd}%
. As a result, this mechanism prevents the actual release of infinite
energy, as it should be in any physically meaningful process: the energy $%
E_{c.m.}$ remains finite in any act of collision although it can be made as
large as one likes.

Now, one can expect that in the situation under discussion the proper time
remains finite since particles are assumed to be usual, without special
fine-tuning. Let us consider, for simplicity, motion in the equatorial plane 
$\theta =\frac{\pi }{2}$. It follows from (\ref{r}), (\ref{z}) that, in the
absence of the turning point, motion between $r_{i}$ and $r_{f}<r_{i}$ takes
the proper time 
\begin{equation}
\tau =m\int_{r_{f}}^{r_{i}}\frac{drN}{\sqrt{A}Z}\text{.}  \label{tau}
\end{equation}

As the particle is taken to be usual, $X>0$ everywhere, $Z>0$ is separated
from zero. Assuming additionally that $\sqrt{A}\sim N$ (like it happens,
say, for the Kerr metric), we see that the integrand in (\ref{tau}) is
finite, so $\tau $ is also finite. When a particle reflects from the turning
point and returns to $r_{0}$, the corresponding time is also finite because
of the same reasonings. Thus in a general model-independent way and without
calculations we can conclude that the proper time before collision is finite.

For the coordinate time $t$, it follows from (\ref{t}) that%
\begin{equation}
t=\int_{r_{f}}^{r_{i}}\frac{drX}{ZN\sqrt{A}}.
\end{equation}

This time is also finite. But, in contrast to $\tau $, the time of travel
between $r_{i}$ and $r_{f}=r_{0}$ grows unbound when $\varepsilon
\rightarrow 0$ in (\ref{e}). Taking $A=N^{2}b^{2}$ where $b$ is some
model-dependent nonzero coefficient, we obtain for time of motion between $%
r_{i}>r_{0}$ and $r_{0}$%
\begin{equation}
t_{0}\approx \frac{\pi }{2b(r_{0})\varepsilon B(r_{0},\frac{\pi }{2})}.
\end{equation}
If one takes into account the time for back motion to $r_{0}$, $t$ acquires
an additional factor 2. Comparison to (\ref{ed}) gives us that 
\begin{equation}
E_{c.m.}\sim t_{0}.
\end{equation}

The contents of the present section agrees with that of Sec.II D\ of \cite%
{rnn} where the particular case of the Reissner-Nordstr\"{o}m metric was
considered.

\section{Case of motion along the axis of symmetry}

There is the case deserving special attention. Let the coordinate $z$ have
the meaning of the polar angle. Let us consider motion along the polar axis,
so $\theta =0$ or $\theta =\pi $. The regularity of the metric near the axis 
$\theta =0$ requires $g_{\phi }\sim \theta ^{2}$. Then, the finiteness of
the term $L^{2}/g_{\phi }$ in (\ref{z}) entails $L=0$, so in (\ref{x}) $X=E$%
. It follows from (\ref{z}) that%
\begin{equation}
Z^{2}=E^{2}-N^{2}m^{2}.  \label{l0}
\end{equation}

Let $N(0,0)=N_{0}$ and $N(\infty ,0)=N_{\infty }$. Then, we can choose any
value of $E$ such that $E<N_{0}$ since this guarantees the presence of the
turning point. If $N_{\infty }<N_{0}$, the particle with an intermediate
energy $N_{\infty }<E<N_{0}$ can fall from infinity. Otherwise, it
oscillates between turning points. Assuming that the presentation (\ref{e})
is valid with $\varepsilon \ll 1$, we obtain the unbound energy $E_{c.m.}$
according to (\ref{la}). This generalizes observation made in \cite{ax} for
the Kerr metric. The same formula (\ref{l0}) applies to the case $\theta =%
\frac{\pi }{2}=const$, $L=0$.

\section{After collision}

Let us consider the scenario described above. One sends particle 1 towards
the centre and, later, particle 2. Particle 1 enters the inner region,
reflects from the potential barrier and collides with particle 2 near $%
r=r_{0}$. As a result, particles 3 and 4 are created. We assume that in the
act of collision both the energy and angular momentum are conserved:%
\begin{equation}
E_{1}+E_{2}=E_{3}+E_{4}\text{,}  \label{en}
\end{equation}%
\begin{equation}
L_{1}+L_{2}=L_{3}+L_{4}\,\text{,}  \label{l}
\end{equation}%
whence%
\begin{equation}
X_{1}+X_{2}=X_{3}+X_{4}\text{.}  \label{x12}
\end{equation}%
We also assume the forward in time condition%
\begin{equation}
X_{i}>0,\text{ }1\leq i\leq 4\text{,}  \label{ft}
\end{equation}%
which follows from (\ref{t}) and $\dot{t}>0$. In contrast to the black hole
case, where $N=0$ on the horizon, now $N>0$ everywhere, so the case $X_{i}=0$
is excluded. Also, we assume that $X_{i}=O(1)$ do not become small near $%
r_{0}$.

The conservation of the radial momentum gives us, according to (\ref{r}),
that

\begin{equation}
Z_{1}-Z_{2}=Z_{4}-Z_{3}\text{.}  \label{z12}
\end{equation}

For small $N$, 
\begin{equation}
Z_{i}\approx X_{i}-\frac{N^{2}}{2X_{i}}\left( \frac{L_{i}^{2}}{g_{\phi }}%
+m_{i}^{2}\right) \text{.}  \label{zn}
\end{equation}%
The main terms give us 
\begin{equation}
X_{1}-X_{2}\approx X_{4}-X_{3}\text{.}  \label{x34}
\end{equation}%
It follows from (\ref{x12}), (\ref{x34}) that 
\begin{equation}
X_{1}\approx X_{4}\text{, }X_{2}\approx X_{3}\text{.}  \label{x3}
\end{equation}%
The main corrections give us from (\ref{z12}), (\ref{zn}) that%
\begin{equation}
\frac{1}{X_{2}}\left( \frac{L_{2}^{2}}{g_{\phi }}+\frac{L_{3}^{2}}{g_{\phi }}%
+m_{2}^{2}+m_{3}^{2}\right) \approx \frac{1}{X_{1}}\left( \frac{L_{1}^{2}}{%
g_{\phi }}+\frac{L_{4}^{2}}{g_{\phi }}+m_{1}^{2}+m_{4}^{2}\right) \text{,}
\end{equation}%
where $g_{\phi }$ is taken in the point of collision $r=r_{0}$.

For fixed $E_{1,2}$ and $L_{1,2}$ (hence, $X_{1}$ and $X_{2}$), we are
interested in the solutions for which $E_{3}$ grows with (\ref{ft}) satisfied%
$.$ According to (\ref{x3}), $X_{3}=X_{2}$ is also fixed, hence this implies
that $L_{3}=\frac{E_{3}-X_{3}}{\omega }$ is large. Let us assume that $%
\omega >0$ everywhere. Our goal can be achieved if, say, $E_{4}\rightarrow
-\infty $, $L_{4}\rightarrow -\infty $, $E_{3}\rightarrow \infty $, $%
L_{3}\rightarrow \infty $. This implies that orbits with large negative
energy do exist. In principle, this is possible even in the absence of the
horizon. Then, insofar as all masses $m_{i}\ll M$ where $M$ is the mass
corresponding to the metric (\ref{met}), there are no bounds on the ratio $%
\frac{E_{3}}{E_{1}+E_{2}}$ on this scale. In this sense, this is the
standard situation for the Penrose process.

\section{Example: regular star-like configurations versus vacuum-like black
holes}

In this section, we give an example of physically relevant objects to which
the scenario of collision under discussion can apply. (Another example based
on the Bardeen spacetime \cite{bard} was given in Sec. IV of \cite{nn1}).\
In a sense, the RN or Kerr naked singularity can be obtained by deformation
of the metric of the corresponding extremal black hole. In a similar way,
the required regular starlike configuration can be obtained by deformation
of a regular extremal black hole. As such an example, we can choose the
regular black hole with the de Sitter core proposed in \cite{dym}. Let us
consider the spherically symmetric metric%
\begin{equation}
ds^{2}=-fdt^{2}+\frac{d\rho ^{2}}{f}+r^{2}(u)d\omega ^{2}\text{.}  \label{f}
\end{equation}

In \cite{dym}, it is assumed that the matter satisfies the vacuum-like
equation of state $p_{r}=-\rho $ ($p_{r}$ is the radial pressure, $\rho $ is
the energy density). Then, $r=u$. However, this is not necessary. We can
consider (\ref{f}) with more general equations of state (which become
vacuum-like near the origin to ensure regularity). We only require (i) the
de Sitter core for small $r$ \cite{dym}, (ii) asymptotic flatness, $%
f\rightarrow 1$ when $u\rightarrow \infty $. As $f=1$ both at infinity and
near the origin $r=0$, it must have a minimum in between. For simplicity, we
assume that there is only one such a minimum. If $f$ has two zeros, we have
a nonextremal black hole, if it has one double zero in the point of minimum,
a black hole becomes extremal, if $f>0$ there is no black hole at all
(starlike configuration). The main purpose of \cite{dym} was to obtain a
regular black hole. By contrary, now we are interested in a starlike
configuration which is close to the extremal black hole in the sense
described above (see eq. \ref{e}). If%
\begin{equation}
f=f_{0}+a(u-u_{0})^{2}
\end{equation}%
near $u=u_{0}$, and $f_{0}\rightarrow 0$, previous general consideration
applies. Then, in the background (\ref{f}) with aforementioned properties,
one can obtain unbound $E_{c.m.}$ for test particles without the horizon or
naked singularity. In the absence of the ergoregion, extraction of energy
does not occur. However, due to large $E_{c.m.}$, creation of superheavy
particles is possible.

\section{Shells}

Instead of test particles, let us consider collision of shells which move in
opposite directions. We can divide the act of collision to the set of
individual collisions of small constituents. For example, if shells are
spherical, natural division consists in collision between particles with the
same value of angle variables. Then, for each pair of colliding elements, we
can apply again eq. (\ref{cm}), (\ref{ga}). The values of $X_{i}$ and $Z_{i}$
can be obtained (for given initial conditions) from equations of motion.
These equations differ from (\ref{t}) - (\ref{z}) due to the effect of
self-gravitation. Say, for collision of charged shells, one obtains (see eq.
74 of \cite{rnn}) that%
\begin{equation}
E_{c.m.}^{2}=2m^{2}+\frac{2m^{2}}{f}(\left\vert \dot{R}_{1}\dot{R}%
_{2}\right\vert +\sqrt{\dot{R}_{1}^{2}+f}\sqrt{\dot{R}_{2}^{2}+f})\text{, }
\end{equation}%
$f\equiv N^{2}$ is taken in the region between shells in the coincidence
limit. The law that governs the dependence $R_{i}(\tau )$ is different for
test particles and constituents of self-gravitation shells and can be
described in terms of different effective potentials. However, the structure
of the expression (\ref{ga}) is universal. Therefore, insofar as $f$ is
small, $E_{c.m.}^{2}$ is large. And, previous explanation based on
presentation (\ref{e}) is still valid. (It is worth stressing that it is
important that shells move in the opposite directions before collision. For
motion in the same direction, the effect of inbound $E_{c.m.}$ is absent 
\cite{sh}.)

Thus inasmuch as we are interested in the effect of gaining unbound $%
E_{c.m.} $ only, there is no need to analyze the whole history of the shell
(which, however, is of interest by itself). What is important is smallness
of $N^{2}$ and the possibility that an inner shell bounces back from some
surface. This can be achieved either due to the naked singularity or the
effective potential barrier of a finite height. Actually, some restrictions
(not related to our subject) on $E_{c.m.}$ come from requirement that
description of shells is macroscopic, so they should contain a large number
of constituents (see Sec. III C of \cite{rnn}).

\section{Discussion and conclusion}

In previous scenarios, the following difficulties were present: (i) the
necessity to ensure fine-tuning, (ii) severe bounds on the energy of
products of collisions measured at infinity, (iii) if collisions are
arranged due to naked singularities, the problem with the cosmic censorship
arises, (iv) if collisions with large $E_{c.m.}$ occur with no horizons or
naked singularities, this requires effects of self-gravity, so for test
particles such collisions could not be realized. Meanwhile, now we see that
all these difficulties can be avoided for regular star-like configurations,
so the effect exists even for test particles. In particular, if there exists
an ergoregion (that is, in principle, is possible even without the horizon -
see some example in \cite{noh}), the collisional Penrose process should
become much more efficient than for the BSW effect. We described an unified
picture of the scenario that ensures the unbound $E_{c.m.}$ in head-on
collisions with the metrics close to forming the horizon but when the
horizon does not form.

Apart from this, the advantage of collisions under discussion consists in
that we can safely neglect the role of gravitational radiation \cite{berti}.
Such radiation bounds the BSW effect since it "spoils" special (critical)
trajectories with fine-tuning of parameters required for this effect.
However, now fine-tuning is not required at all, so small perturbation due
to an additional force do not change the whole picture qualitatively.

It is instructive to classify main types of the effect under discussion -
see Table 1.

\begin{tabular}{|l|l|l|l|l|l|}
\hline
Relevant references & Relative direction & Horizon & Naked singularity & 
Fine-tuning & Self-gravity \\ \hline
\cite{psk}, \cite{ps} & $-$ & $+$ & $-$ & $-$ & $-$ \\ \hline
\cite{ban} & $+$ & $+$ & $-$ & $+$ & $-$ \\ \hline
\cite{acqbh} & $+$ & $-$ & $-$ & $-$ & $-$ \\ \hline
\cite{kerr}, \cite{rnn}, \cite{ax} & $-$ & $-$ & $+$ & $-$ & $-$ \\ \hline
\cite{4} & $-$ & $-$ & $-$ & $-$ & $+$ \\ \hline
present paper & $-$ & $-$ & $-$ & $-$ & $-$ \\ \hline
\end{tabular}

Table 1. Different types of high energy processes with horizons or would-be
horizons.

For shortness, in this table, "self-gravity" means "necessity of
self-gravity to have unbound $E_{c.m.}$" (collision of shells), etc.

Thus the present kind of high energy collision in a strong gravitational
field looks promising since it relaxes or weakens strong restrictions
typical of the BSW\ effect required for observation (at least in principle).
In doing so, only the vicinity of the would-be horizon is important in
accordance with the spirit of black hole physics (even in the absence of a
black hole!). In this sense, in all effects described in Table 1, it is
necessary that a system posses either the true horizon or time-like surface
which in a sense is close it.

\end{document}